\documentclass[twocolumn,epjc3]{svjour3}  
\smartqed  
\RequirePackage{graphicx}
\usepackage[utf8]{inputenc}
\usepackage{hyperref}\RequirePackage{fix-cm}
\usepackage[rightcaption]{sidecap}
\usepackage{pdfpages}
\graphicspath{ {images/} }

\newcommand{\TabBesBeg}[1][1.0]{ %
\let\MyTable\table
\let\MyEndtable\endtable
\renewenvironment{table}[1]{\begin{SCtable}[#1]##1}{\end{SCtable}}}
\newcommand{\TabBesEnd}{ %
\let\table\MyTable
\let\endtable\MyEndtable}
\newcommand{\FigBesBeg}[1][1.0]{ %
\let\MyFigure\figure
\let\MyEndfigure\endfigure
\renewenvironment{figure}[1]{\begin{SCfigure}[#1]##1}{\end{SCfigure}}}
\newcommand{\FigBesEnd}{ %
\let\figure\MyFigure
\let\endfigure\MyEndfigure}

\usepackage{float}
\usepackage[utf8]{inputenc} 
\usepackage{braket}
\usepackage{tikz}
\usetikzlibrary{quantikz2}
\usepackage{amssymb}
\usepackage{graphicx} 
\usepackage{bbm}
\usepackage{soul}

\journalname{Eur. Phys. J. A}

\begin{document}

\title{Triply-heavy/strange baryons with Cornell potential\\ on a quantum computer }
\titlerunning{Heavy baryons on a quantum computer}  

\author{Nicol\'as Mart\'{\i}nez de Arenaza\thanksref{e1,addr1} 
        \and
        Juan J. G\'alvez-Viruet\thanksref{e2,addr1}
         \and
        and \\ Felipe J. Llanes-Estrada\thanksref{e3,addr1} 
}
\thankstext{e1}{e-mail: nicoma04@ucm.es}
\thankstext{e2}{e-mail: juagalve@ucm.es}
\thankstext{e3}{e-mail: fllanes@fis.ucm.es  Supported by grant PID2022-137003NB-I00 of the Spanish MCIN/AEI
/10.13039/501100011033/; EU’s 824093 (STRONG2020); and Universidad Complutense de Madrid under research group 910309 and the IPARCOS institute.}

\institute{Dept. F\'{\i}sica Te\'orica \& IPARCOS, Fac. CC. F\'{\i}sicas, Plaza de las Ciencias 1, Universidad Complutense de Madrid, 28040 Madrid, Spain. \label{addr1}
}

\date{Received: July 10th 2024 / Accepted: date}

\maketitle

\begin{abstract}
We present a computation of triply-heavy baryons on a quantum computer, employing the Cornell quark model in line with the earlier quarkonium work of Gallimore and Liao.
These baryons are some of the most interesting Standard Model particles which have not yet been detected, as they bear on the short range (colour) behaviour of the nuclear force.
The spectrum here obtained is compatible with predictions from earlier works, with our uncertainty dominated by traditional few-body approximations (size of the variational basis, center of mass recoil, parameter estimation...) and not by the statistical error from 
the quantum computer (deployed here as a small diagonalizer), which turns out to be negligible respect to the other sources of uncertainty, at least in the present unsophisticated few-body approximation. 
We have also substituted one or more heavy quarks for strange quarks.
\end{abstract}

\section{Introduction}

Exploratory use of quantum computer has been increasing~\cite{Seth_Lloyd} in several subfields of quantum physics, such as atomic~\cite{Martinez_2016} or particle physics~\cite{barata_2021,Jordan_2012,Fornetti:2024uel}, as well as in physical chemistry~\cite{Whitfield_2011}.
Not long ago, Gallimore and Liao~\cite{gallimore-2023} employed a few-qubit quantum computer as a small diagonalizer demonstrating its usage for $c\bar{c}$ quarkonium states in the quark model. We here quickly reproduce that calculation, which we use to fix the parameters of the Cornell potential in this context (section~\ref{sec:H}), and then proceed to our object of study, triply-heavy baryons composed of heavy $c$ and $b$ quarks in various combinations. 
These are of particular interest to help disentangle forces at different distances in nuclear physics. Because the ground state radii of these somewhat Coulombic systems scale with the mass $m$ and coupling constant $\alpha_s$ as $1/(m\alpha_s)$,  these baryons are more compact than ordinary ones. Thus, the proportion of meson-exchange and colour-exchange forces is different from ordinary nuclei-forming baryons. Numerous calculations of their spectrum exist, {\it e.g.} \cite{Gomez-Rocha:2023jfr,Najjar:2024deh,Yang:2019lsg,Wei:2016jyk}, including a variational estimate~\cite{Llanes-Estrada:2011gwu} in the effective theory potential Nonrelativistic QCD (pNRQCD) and others which will be quoted below. At least the triply charmed $\Omega_{ccc}$ should be within range of discovery for $e^-e^+$ factories such as Belle-II~\cite{Drutskoy:2012gt} at the top of its energy reach,
and some of them could probably be discovered at the HL-LHC.

In consequence, we dedicate section~\ref{sec_3} to the triply charmed $\Omega_{ccc}$.  In subsection \ref{sec_3.1} we analyze the Cornell Hamiltonian for this baryon, then implement the Hamiltonian on a quantum computer as reported in subsection \ref{sec_3.2}, and analyze the obtained mass in subsection~\ref{sec_3.3} .

Moving onto baryons of mixed $b$ and $c$ flavours, we need a slight modification in encoding the Hamiltonian. After adjusting the mass of the $b$ quark in subsection~\ref{4.2}, we can predict the mixed heavy baryon spectrum in subsection~\ref{sec_4.3}.

Finally we extend the computation to baryons composed of both heavy and strange quarks.
For this we compute Gell-Mann's $\Omega_{sss}$ mass in subsection~\ref{sec_5.1} and use it to fix the $s$-quark mass, so that the rest of section~\ref{sec_5} can be dedicated to mixed strange-heavy baryons, a couple of which have already been hinted at in experiment. Although this is not a strictly nonrelativistic system, it partially anchors the calculation to data and suggests that there is a working chance that the LHC will indeed discover the all-heavy baryons.

Even as the usage of the quantum computer does not presently bring any advantage over computations which can be performed, as they have been for decades, on traditional systems, it is interesting to see many existing calculations reproduced with these machines which hold promise to revolutionize our field when deployed at a larger scale.

\section{Hamiltonian model}\label{sec:H}

We adopt the Cornell static, central potential among nonrelativistic sources~\cite{Griffiths:2008zz} with  $\alpha_s$ the QCD coupling and $\sigma$ the infrared confining string tension. For the effective one-body problem for the relative particle in a two-body, e.g. $c\bar{c}$ system, the Hamiltonian is then
\begin{equation}
H= 2m_c+T+V=2m_c-\frac{\nabla^{2}}{2\mu} +\sigma r -\frac{\alpha_s}{r} . \label{H}
\end{equation}
The reduced mass $\mu=m_c/2$, as well as the potential parameters $\alpha_s$ and $\sqrt{\sigma} $ can be read off table~\ref{tabla_4}.
The eigenvalues of $H$, a central potential, need to be compared to the spin-averaged charmonia; in an $s$-wave, these have quantum numbers $J^{PC}$ given by  1$^{--}$ (the $J/\psi$) and 0$^{-+}$ (the $\eta_c$).

The short-distance strong coupling $\alpha_s$ depends on the gluon scale, and thus on the characteristic overall scale of the state, which we take as the quark mass. We limit ourselves to one-loop in the QCD perturbative expansion
\begin{equation} 
\alpha_s(Q^2)=\frac{4\pi}{\beta_0 \ln(\frac{Q^2}{\varLambda^2})}, \label{runningalpha}
\end{equation}
Eq.~(\ref{runningalpha}) shows the strong, then divergent nature of $\alpha_s$ 
at low momentum transfers, so that light quark computations are affected by larger uncertainties, as will be seen in Section~\ref{sec_5}. The first beta-function coefficient 
\begin{equation} 
\beta_0=11-\frac{2}{3}n_f\label{eq_29}
\end{equation}
depends on the number of active quark flavors  $n_f$ at the scale $Q$. For the heavy quarks in this work, this is taken to be four. One can compute the QCD scale according to
\begin{equation} 
\varLambda^2=m^2\exp\Big(\frac{-4\pi}{\beta_0\alpha_s(m^2)}\Big)\label{eq_30}
\end{equation}
where, for the charm sector, $m=m_c$ is the quark mass from table~\ref{tabla_5a}) and  $\alpha_s(m^2)$ can be found in table~ \ref{tabla_4}.

\begin{figure}
    \centering
    \includegraphics[scale=0.20]{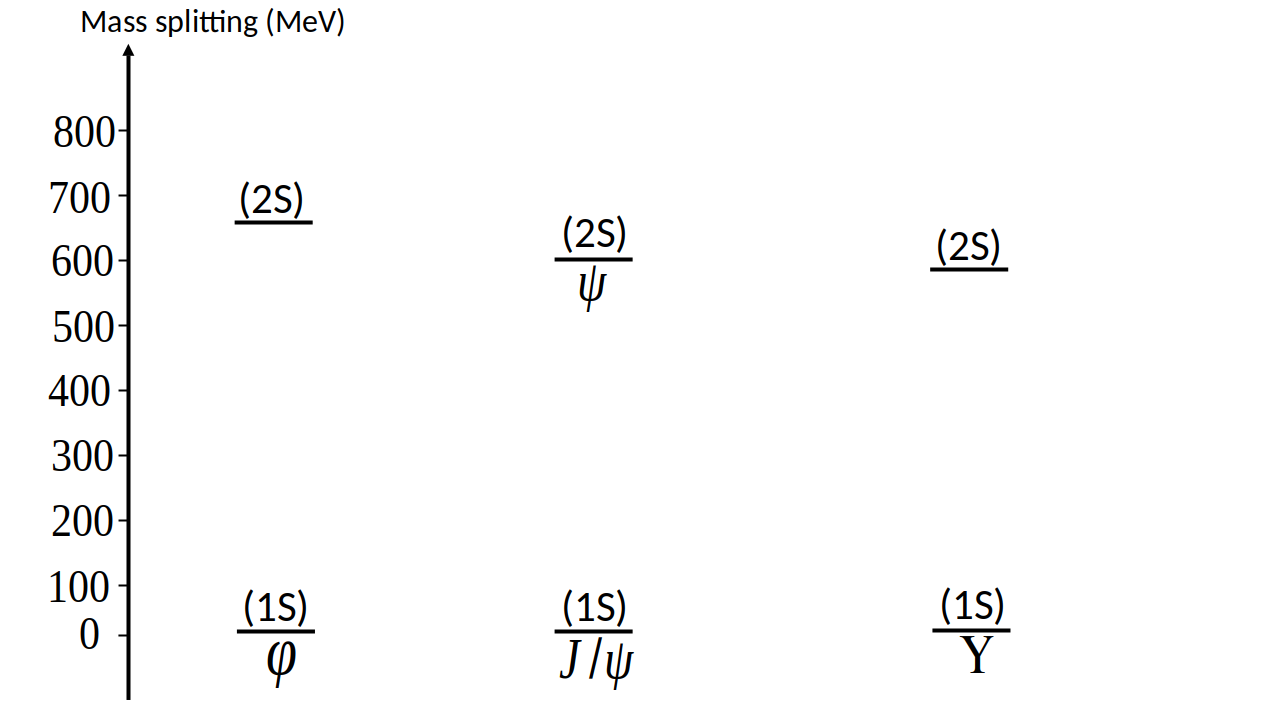}
    \caption{\label{feee}The energy differences between $1s$ and $2s$ states (with $J^P=1^-$) are shown to be very weakly dependent on flavour, suggesting an also independent string tension $\sigma$.}
    
\end{figure}

The string tension $\sigma$ is taken as flavour and scale independent as it represents long-distance physics (confinement) dominated by the representation of the colour sources and gluodynamics~\cite{Greensite:2011zz}.
We remind the reader of a straightforward empirical piece of support for this statement in figure~\ref{feee}:
the energy difference between the $1s$ fundamental (Coulombic, at least for heavy quarks) and $2s$ first excited (feeling the confinement potential already) states happens to be rather independent of the quark flavour.
That congruence of the levels for different flavours indeed suggests that the confining interaction is flavour-independent.

Although the particle number is fixed (but see~\cite{Galvez-Viruet:2024hry} for a number-changing treatment), we can adopt the formalism of second quantization for the filling number of each of the variational basis states, so that $H$ is written in terms of standard fermion-pair creation and destruction operators,
\begin{equation} 
 H_N= \sum_{n',n=0}^{N-1} \braket{n'|T+V|n}a_{n'}^\dagger a_n\ . \label{eq_3}
\end{equation}
The states of the basis, labeled by $\ket{n}$, correspond to the $s$-wave eigenstates of the harmonic oscillator.
Because this is not a basis of eigenstates of the Cornell Hamiltonian, unless the limit  $N \rightarrow \infty$ is taken, we are adopting a variational approximation. In this work, $N$ is rather low as the purpose is to try the computation out on a quantum computer (but this variational error is systematically improvable by increasing the basis size).
The variational theorem of Hylleraas-Undheim-MacDonald ~\cite{hyle},~\cite{Mc}, indicates that the eigenvalue of the truncated Hamiltonian $H_N$ will bound the exact mass from above.

The occupation of these states is naturally written as $\ket{f_{N-1}...f_1f_0}$ where $f_n$=0,1 decides the presence or absence of the $c\bar{c}$ pair in state $\ket{n}$. The usability of this formulation as regards a quantum computer is evident, as each 
$f_n$ can be simply represented by one qubit in the states
 $\ket{0}= (1,0)^T$, $\ket{1}= (0,1)^T$.   

The coordinate representation of the $l=0$ states in terms of the spherical coordinate $r$ is
\begin{equation}
\braket{r|n} = (-1)^n\sqrt{\frac{2n!}{d^3 \Gamma \left(n+\frac{3}{2}\right)}}\exp(-\frac{r^2}{2d^2})L_n^{1/2}(\frac{r^2}{d^2}) \label{Laguerre}
\end{equation}
in terms of the generalized Laguerre~\cite{La} polynomials $L_n^{1/2}$ (their orthogonality with the adequate measure has been analytically as well as numerically checked). 
The constant $d$ is the oscillator's characteristic length  and it is taken as a variational parameter. 

Evaluating the potential  in Eq.~(\ref{H}) requires the matrix elements
\begin{eqnarray} 
 \braket{n'|r|n}=(-1)^{n'+n}\frac{4d}{\pi(1-4n^2)} \nonumber \\ \sqrt{\frac{\Gamma\left(n'+\frac{3}{2}\right)\Gamma\left(n+\frac{3}{2}\right)}{{n'}!n!}}\enskip \cdot _2F_1\left(2,-n';\frac{3}{2}-n;1\right)\label{eq_5},
\end{eqnarray}
and
\begin{eqnarray} 
 \braket{n'|r^ {-1}|n}=(-1)^{n'+n}\frac{4d^{-1}}{\pi(1+2n)}\nonumber \\
\sqrt{\frac{\Gamma(n'+\frac{3}{2})\Gamma(n+\frac{3}{2})}{n'!n!}} \enskip \cdot _ 3F_2(\frac{1}{2},1,-n';\frac{3}{2},\frac{1}{2}-n;1). \label{eq_6}
\end{eqnarray}
In these two equations, standard hypergeometric functions have been used,
\begin{eqnarray} 
  _2F_1(a,b;c;z)= \sum_{k=0}^{\infty}\frac{abz^k}{ck!} \nonumber \\
_3F_2(a_1,a_2,a_3;b_1,b_2;z)= \sum_{k=0}^{\infty}\frac{a_1a_2a_3z^k}{b_1b_2k!} \ ,\label{eq_8}
\end{eqnarray}
where $z\leq 1$.

The kinetic energy represented in the oscillator basis (which diagonalizes an interacting theory, not the free one and thus not the kinetic energy) reads
\begin{eqnarray}
\langle n'| T| n\rangle
=\frac{\omega}{2}\left[\left(2n+\frac{3}{2}\right)\delta_{nn'}\right. \nonumber \\ -\left.\delta_{n'+1,n}\sqrt{n\left(n+\frac{1}{2}\right)}      -\delta_{n'-1,n}\sqrt{(n+1)\left(n+\frac{3}{2}\right)} \right] \nonumber \\ \label{T_matrix_element}
\end{eqnarray}
with angular frequency
$\omega$ defined by $\omega=\frac{1}{\mu d^2}$, whose values, together with those of $d$, are pushed to table~\ref{tabla_4}.

\newpage
\section{Charmonium\label{sec:charmonium}}

\subsection{Representation of $H$ by means of quantum gates
\label{sec_2.2}}
In this subsection we render $H$ in terms of the quantum gates which have been implemented on the quantum computer.

First, the creation and destruction operators of Eq.~(\ref{eq_3}) 
can be translated, while maintaining $[a_n,a_n^\dagger]=0$, through the Jordan-Wigner transform~\cite{jordannn} on an 
$n+1$ qubit circuit, 
\begin{equation} \label{jordan}
 a_n^\dagger= \frac{1}{2}\Bigr(\prod_{j=0}^{n-1}\sigma^z_j\Bigr) (\sigma^x_n-i\sigma^y_n), \enskip a_n =\frac{1}{2}\Bigr(\prod_{j=0}^{n-1}\sigma^z_j\Bigr) (\sigma^x_n+i\sigma^y_n) ,
\end{equation}
in terms of Pauli gates which act on the $n$th qubit.

To reach $1\%$ uncertainty for the charmonium ground state, 
a three-vector basis with $N=3$  ($n=0,1,2$) suffices
and $H$ in Eq.~(\ref{eq_3}) becomes a  $3\times 3$ matrix 
\begin{equation} 
H_3=\sum_{i=0}^9 H_3^i\ .
\end{equation}
Each of the $H_3^i$ terms is a one-qubit gate readily implementable on a quantum computer, and they are expressed in terms of the $\sigma^j_n$ in the appendix, in table~\ref{tabla_1}, with their corresponding numerical values with the parameters of $c\bar{c}$ in table~\ref{tabla_8}.

\subsection{Variational basis \label{subsec:variational}}
We here choose a reasonable set of states to deploy the Rayleigh-Ritz~\cite{Mc} variational method which indicates that, given the state vector $\ket{\psi(\theta_1,\theta_2)}$ 
and the Hermitian observable $\mathcal{O}$,  the smallest eigenvalue $\epsilon_0$ of $\mathcal{O}$ can be bound above by the expectation value
\begin{equation}
    \braket{\mathcal{O}}_\psi=\braket{{\psi(\theta_1,\theta_2)}|\mathcal{O}|\psi(\theta_1,\theta_2)} \geq \epsilon_0\ .
\end{equation}
The optimal bound is obtained by varying the parameters
$\theta_1$ and $\theta_2$ looking for the minimum ${\rm min}(\braket{{H_3}}_\psi)$ over this wavefunction family.

Since we employ a three-orbital basis to assign the $c\bar{c}$ pair, the variational wavefunction is conveniently  constructed as
$\bra{001}U(\theta_1,\theta_2)$ in terms of a unitary operator  
\begin{equation}
    U(\theta_1,\theta_2)=\exp\Bigr(\theta_1(a_1^\dagger a_0-a_0^\dagger a_1)+\theta_2(a_2^\dagger a_0 -a_0^\dagger a_2)  \Bigr)\ . \label{eq_14}
\end{equation}
Substituting Eq.~(\ref{jordan}) we can reduce the exponential to a polynomial of Pauli $\sigma$ matrices,
\begin{eqnarray}
U(\theta_1,\theta_2)= \nonumber \\
\exp\Bigr(\theta_1 \frac{i}{2}( \sigma^x_1\sigma^y_2-\sigma^y_1\sigma^x_2   )+\theta_2 \frac{i}{2}(\sigma^x_0\sigma^z_1\sigma^y_2-\sigma^y_0\sigma^x_1\sigma^x_2) \Bigr).\nonumber \\
\end{eqnarray}

Because of the structure of Eq.~(\ref{eq_14}), which does not change the number of $c\bar{c}$ pairs, the set of linear combinations of the three vectors with exactly one $c\bar{c}$ pair in each possible configuration,
\begin{equation}
\ket{f_1f_2f_3}\in {\rm span}\{\ket{100}, \ket{010},\ket{001}\}
\end{equation}
is an invariant subspace of
$\log(U(\theta_1,\theta_2))$ which allows to represent $U$ in that subspace with matrix
\begin{equation}
log(U(\theta_1,\theta_2))=
\begin{pmatrix}
0 & 0 & \theta_2\\
0 & 0 & \theta_1\\
-\theta_2 & -\theta_1 & 0 \\
\end{pmatrix} ,
\end{equation}
with the columns corresponding, from left to right, to $\ket{100}$, $\ket{010}$ and $\ket{001}$.

The eigenvalues of this matrix are  $-i\alpha$, $i\alpha$ and 0, with $\alpha=\sqrt{\theta_1^2+\theta_2 ^2}$.  Such diagonal form allows for easy exponentiation, so that $U$ becomes represented by
\begin{eqnarray}
U(\theta_1,\theta_2)=
\begin{pmatrix}
c^2\beta c\alpha +s^2\beta & s\beta c\beta(c\alpha-1) & -c\beta s\alpha\\
s\beta c\beta(c\alpha-1)& c^2\beta c\alpha +s^2\beta & -s\beta s\alpha\\
s\alpha c\beta & s\alpha s\beta & c\alpha \label{eq_16a}
\end{pmatrix} \ ,
\end{eqnarray}
where the convenient variable change $s\beta={\rm sin}\beta\equiv \theta_1/\alpha$ has been introduced.
For example, the application of this operator to the third brac of the (conjugate) basis produces the state
\begin{eqnarray}
\ket{\psi(\alpha,\beta)}= \bra{001}U(\theta_1,\theta_2) \nonumber \\ =\sin\alpha \cos\beta \ket{100}+\sin\alpha \sin\beta \ket{010}+\cos\alpha\ket{001}\ .\nonumber \\ \label{eq_16}
\end{eqnarray}

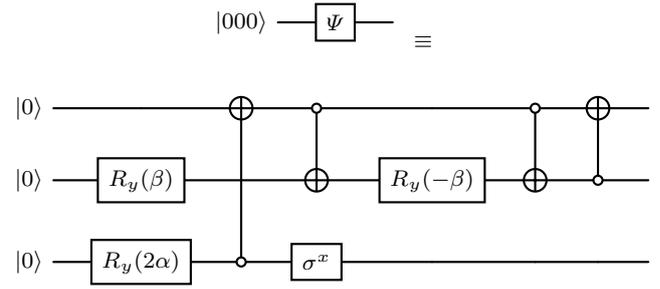
\begin{figure}
\begin{center}
    \begin{quantikz}
    \centering
\lstick{$\ket{000}$}&   \gate{ \Psi} &  \\
\end{quantikz}
$ \enskip\equiv \enskip$
    \begin{quantikz}
    \centering
\lstick{$\ket{0}$}& &  \targ{}   & \octrl{1}  &  & \octrl{1} & \targ{}&\\
\lstick{$\ket{0}$}&  \gate{R_y(\beta)} & & \targ{} & \gate{R_y(-\beta)}   & \targ{} & \octrl{-1}&\\ 
\lstick{$\ket{0}$}& \gate{R_y(2\alpha)} & \octrl{-2} & \gate{\sigma^x} & &&&\\
\end{quantikz}
\end{center}
   \caption{ Quantum computer circuit initializing the state of Eq.~(\ref{eq_16}) by means of unitary gates once the variational parameters $\alpha$, $\beta$ are fixed. \label{fig_1} }
\end{figure}

Therefore, the three-qubit system is elegantly represented by the two coordinates over a sphere of unit radius.
 Such state $\ket{\psi(\alpha,\beta)}$ may be prepared, employing only unitary operators on the circuit as shown in Figure~\ref{fig_1}.

Other gates which appear in the circuits there depicted are standard~\cite{Nielsen_Chuang_2010,Ee}, namely $R_y(\alpha)$ (a $\alpha/2$ rotation around $OY$);
$CNOT$, represented by $\oplus$ 
on the qubit over which it acts, the target, and a white dot over the controlling qubit;
and the Hadamard gate, represented in figure~\ref{fig_1a} by $H$, and that
takes $\ket{0}\to\frac{\ket{0}+\ket{1}}{\sqrt{2}}\equiv\ket{+}$ and $\ket{1}\to \frac{\ket{0}-\ket{1}}{\sqrt{2}}\equiv\ket{-}$.

\subsection{Measuring $\langle H\rangle_\psi$ on the quantum computer \label{sec.2_4}}
The measurement of the Hamiltonian's expected value is obtained by applying $H_3$ 
to the state $\ket{\psi(\alpha,\beta)}$ prepared by the circuit in figure~\ref{fig_1}.

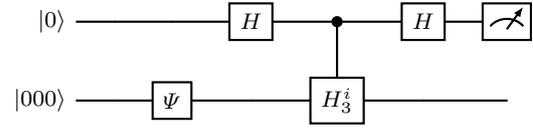
\begin{figure}
\begin{center}
    \begin{quantikz}
    \centering
\lstick{$\ket{0}$}& &  &\gate{H}  &\ctrl{1}&\gate{H}&\meter{}\\
\lstick{$\ket{000}$}& &  \gate{ \Psi}&  &\gate{H^i_3}&& \\
 \end{quantikz}
\end{center}
   \caption{Quantum computer circuit to extract the expectation value of the Hamiltonian (represented by the matrix elements in table
 \ref{tabla_1}) on the state prepared by the circuit in figure~\ref{fig_1}. (Note that $H$ acting on the uppermost qubit is the Hadamard gate and not the Hamiltonian.) \label{fig_1a}} 
\end{figure}

In these first lines we describe the application of the program to run on a classical simulator too. Here, it is convenient to add an ancillary qubit to directly obtain the expected value, so a second application of the operator generating the state $\ket{\psi(\alpha,\beta)}$ becomes unnecessary. Figure~\ref{fig_1a} shows the circuit actually employed.

Each  $H_3^i$ gate therein corresponds to a piece of the Hamiltonian (see table \ref{tabla_1}), written in terms of the Pauli matrices ($\sigma^x_i$, $\sigma^y_i$, $\sigma^z_i$) on the $i$th qubit. 
Each  $H_3^i$ gate is applied to the state $\Psi$,
followed by a {\tt CNOT} controlled by the auxiliary qubit, resulting in its entanglement with the state to be measured. 

At the rightmost end of figure~\ref{fig_1a}   a measurement over the auxiliary qubit is applied.
The result of the measurement is then multiplied by 
the weight (accompanying the Pauli matrix)  corresponding to each of the $H_3^i$ (which we read off table~\ref{tabla_1}).

Now let us proceed to the actual quantum computer. 
We have run the computation in the IBM Eagle chip. 
Due to the cross-shaped disposition of the qubits, 
4-qubit operations suffered from significant noise\footnote{See $https://quantum.ibm.com/services/resources?system=ibm \_osaka$}, particularly among qubits which are at the end of different arms of the cross.

As a mitigating solution, we have opted for not using the additional ancillary qubit (running on three qubits allows to position them on a line). Therefore we need to compute $\braket{\psi(\alpha,\beta)|H_3|\psi(\alpha,\beta)}$ by applying $H_i$ to the state $\ket{\psi(\alpha,\beta)}$, measuring on the circuit, and then projecting again over $\ket{\psi(\alpha,\beta)}$, paying once more the state-preparation gates but avoiding the additional qubit.

\subsection{Numerical $c\bar{c}$ ground state masses} \label{resultadose}

\begin{figure}[H]
    \centering
    \includegraphics[scale=0.5]{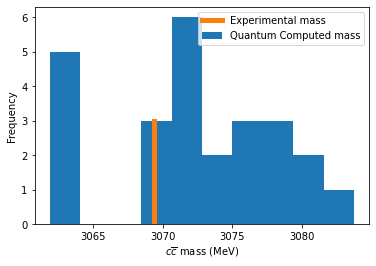}
    \caption{\label{histograma}
    Ground state $c\bar{c}$ mass extracted from 25 runs on an IBM Eagle-chip quantum computer with the Cornell model. 
}
\end{figure}

In a first pass we employed a classical simulator as a precomputer, to explore the landscape of the  $(\alpha,\beta)$ plane
to locate the region where $\langle H\rangle$ could be minimum. This happens around $\alpha\simeq 1.45$ and $\beta\simeq 0.15 $ radians.
The expectation value $\braket{H_3}_\psi$ is then computed with several ``shots'' or runs, typically 25, on the actual quantum device. The various calculations are spread over a statistical distribution, whose histogram can be seen in figure~\ref{histograma}. (Analogous plots can be produced for other calculations in this article, but we will not render them as they contain similar information, and only quote the uncertainty on the computed mass in the main text.)
Quoting a 2$\sigma$ uncertainty, the binding energy for spin-averaged charmonium is then extracted as $492\pm 12$ MeV. From this we can obtain the meson mass by adding the quark masses,
 $M=2M_c+\braket{H_3}_\psi$.
 
 This leads to a computed $M=3072\pm 12$ MeV which needs to be compared to the spin-averaged mass between the vector $J/\psi$ and pseudoscalar $\eta_c$ masses~\cite{Review},
 \begin{eqnarray}
\overline{M} =\frac{3M_{1^-}^{\rm exp}+M_{0^-}^{\rm exp}} {4} \nonumber \\ 
=\frac{3\cdot3097.900 \pm 0.006+2983.9 \pm 0.4}{4} {\rm MeV}
\nonumber \\ 
= 3069.4\pm0.1 {\rm MeV}
\ .
\end{eqnarray}
The corresponding wavefunction components of  $\ket{\psi}_{1s}$ are listed on table~\ref{tabla_10} in the appendix.

We can then compute the first excited state, $\ket{\psi_{2s}}$. 
For this need to impose orthogonality with  $\ket{\psi_{1s}}$
to reduce contamination of the ground state when searching for the new energy minimum. 
We again employ a simulator to quickly scan the possible values of $\alpha$ and $\beta$, which are found to be approximately 5 and 4.9 radians respectively. 

Upon measuring ten times on the quantum computer with those two values, the Hamiltonian for the first excited state, $\braket{H_3}_\psi$, yields 1199$\pm$10MeV.

Adding the quark masses, we obtain a first excited state at 3779$\pm$10MeV, which is somewhat larger than the experimental spin-average of the first excited states,
 $\psi(3686.10\pm0.06)$ (for $J=1$) and $\eta _c^{2s}$ ($3637.7\pm$1.1) (for $J=0$, 
 which returns a value of $\overline{M}_{2s}=3674.0\pm 0.3$ MeV. This is a  2.9\% error, likely stemming from the variational treatment in a small space, upon having truncated Eq.~(\ref{eq_3}) at N=3, and could easily be reabsorbed in the model's string tension $\sigma$ if wished.
 The wavefunction $\ket{\psi_{2s}}$ of this first state is also quoted in table~\ref{tabla_10}.

Once the $\sigma$ and $m_c$ parameters have been shown to be reasonably adjusted to these two data, we may proceed to produce predictions for a triply heavy quark system, to which we dedicate the next section.

\section{Triply charmed baryons in an $s$-wave \label{sec_3}}
\subsection{Three-body Hamiltonian\label{sec_3.1}}

We now turn to three heavy quarks, all in an $l=0$ state
(a system analogous to Fermi's $\Delta$ baryon resonance but with all heavy quarks). The resulting baryon,
usually named $\Omega_{ccc}$, has of course been widely studied in the past~\cite{Llanes-Estrada:2011gwu} and is a good benchmark for new theoretical methods, although it has not been experimentally detected yet.

The interaction in the Cornell model is in the triangle configuration (as opposed to the star configuration~\cite{Dmitrasinovic:2013jta,Salom:2021rmr}), with potential energy 
$ V(r)=V_{12}+V_{13}+V_{23}$ which is a sum of three terms, each one leaving a spectator quark, with the pairwise potential
 $V_{ij}$ of Eq.~(\ref{H}) substituting $r$ by $|r_i-r_j|$ ($r_i$, with $i=1,2,3$, is each quark's coordinate respect to an arbitrary origin)
\begin{equation}
\begin{array}{cc}
 V(r)=-\frac{\alpha_s}{2}\Bigr(\frac{1}{|r_1-r_2|}+\frac{1}{|r_1-r_3|}+\frac{1}{|r_2-r_3|}\Bigr) 
      &  \\+ \frac{\sigma}{2} \Bigr(|r_1-r_2|+|r_1-r_3|+|r_2-r_3|\Bigr).
      & 
 \label{eq_19}
\end{array}
\end{equation}
Note that this equation has an explicit $\frac{1}{2}$ factor not in Eq.~(\ref{H}). This is because the colour factor for a $qq$ pair in a meson is computed from $\frac{1}{3}$T$_{ij}^a$T$_{ji}^a=4/3$ (with  $T_{ij}^a$ the colour $SU(3)$ generator), in a baryon on the other hand each colour factor takes the form
\begin{equation}
\braket{\psi|V_{12}|\psi}_{colour}=-\frac{1}{6} \epsilon^{ijk}\epsilon^{lmn}
T_{il}^aT_{jm}^a\delta_{kn}
 =\frac{2}{3}\ ,
\end{equation}
 and we have chosen not to rescale the constants, which include (for convenience) the 4/3 factor. 

Other interactions proposed in the literature, such as those mediated by Goldstone-boson exchanged
among constituent quarks~\cite{Ronniger:2011td}, are expected to be less important for heavy baryons with their more compact scales (but see~\cite{Day:2012qf} for a computation), so we remain within colour-based interactions in this work.

The kinetic energy in turn takes the form
\begin{equation}
T= T_1+T_2+T_3-T_{cm}\label{eq_21.a}.
\end{equation}
$T_i$ is here each of the quarks kinetic energies, and  $T_{cm}$ that of the center of mass.
 $T_i$ is computed by means of Eq.~(\ref{T_matrix_element}). The variational parameter  $d =\frac{1}{\sqrt{\mu \omega}}$, \textit{c.f.} Eq~(\ref{Laguerre}), is selected by taking $\omega=$ 562.9 MeV, \textit{c.f.} Eq~(\ref{T_matrix_element}), independently of the selected quark.
 The subtraction of $T_{cm}$ is due to having referred each momentum to a fixed origin instead of employing Jacobi coordinates separating the center of mass: the sum of the three quark momentum vectors is not zero, $\sum {\bf p}_i\neq {\bf 0}$. Hence, there is a spurious contribution to the kinetic energy due to the center of mass motion.

 In the philosophy of maintaining the correction to the one-body energy at the one-body level, we will approximate the subtraction of the kinetic energy as proposed in the literature~\cite{DEGREGORIO2021136636} by noting that the center of mass recoil decreases as the inverse of the constituent number, $1/A$. A reasonable approximation is then to write down 
 \begin{equation}
   T_{cm}\approx \sum_{i=0}^3 T_i/A  \ ,
 \end{equation} that is, discounting a part of the kinetic energy of each of the constituents. The terms thereby corrected do naturally appear as the diagonal terms of
\begin{equation}
\begin{array}{cc}
T_{cm} =\frac{p_{cm}^2}{2M}=\frac{(p_1+p_2+p_3)^2}{2M} \\ = \frac{(p_1^2+p_2^2+p_3^2+2 p_1\cdot p_2+2 p_1\cdot p_3+2p_2\cdot p_3)}{2M} \thickapprox \frac{p_1^2+p_2^2+p_3^2}{2M} \\
=\sum_{i=0}^3\frac{m_i}{M} T_i \label{eq_21.c}
\end{array}
\end{equation}
The total kinetic energy may then be written as
\begin{equation}
T= \sum_{i=0}^3\frac{M-m_i}{M} T_i\label{eq_21.b}.
\end{equation}
or, for three equally massive quarks,
\begin{equation}
T= \frac{2}{3}\sum_{i=0}^3 T_i\label{eq_21.bb}.
\end{equation}
 Those neglected are the crossed-terms of the Laplacian, which break the one-body philosophy. The correction does improve for large numbers of particles, falling as $1/A$, but for a three-body system  it introduces a sizeable uncertainty. A future improvement to this investigation is to clearly adopt Jacobi coordinates separating the CM. For the time being, we need to assign to this correction a 50\%  uncertainty of its value, reflecting the approximate treatment, and this comes to dominate the uncertainty band.

\subsection{Representation of the three-body Hamiltonian by quantum gates \label{sec_3.2}}
The basis $\ket{n_i}$ of Eq.~(\ref{eq_3}) needs to be redefined for a three-quark system. Once having sacrificed some of the accuracy for simplicity, we can define separate orbitals for each of the quarks, employing Eq.~(\ref{Laguerre}) with an additional particle index $j$,
\begin{equation}
\begin{split}
    &\braket{r_j|n_i} =\\& (-1)^{n_i}\sqrt{\frac{2n_i!}{d_j^3 \Gamma (n_i+3/2)}} exp(-\frac{r_j^2}{2d_j^2})L_{n_i}^{1/2}(\frac{r_j^2}{d_j^2})\ . \label{eq_22}
\end{split}
\end{equation}

Here $r_j$ is each quark's visual to the arbitrary origin.
The three-particle basis will be described by three quantum numbers  $\ket{k}= \ket{n_1n_2n_3}$, with $(n_1,n_2,n_3)$ 
labelling the quark orbitals from Eq.~(\ref{eq_22}) 
($n_i=n_i(r)$ is thus a shorthand for $\braket{r_j|n_i}$).
The same indexing extends to the variational parameter $d_j$ that is related to the kinetic energy scale $\omega$ by $d_j=(m_j\omega)^{-1/2}$ in terms of each quark's mass.

We wish to discuss the ground and the first two excited states
 $\ket{k}$=$\ket{\hat{0}}$, $\ket{\hat{1}}$ and $\ket{\hat{2}}$ of the Hamiltonian in Eq.~(\ref{eq_3}) with $N=3$ (the caret $\hat{}$ is being used to avoid confusion with Eq.~(\ref{eq_16}) ), and we also replace the indices $n'$, $n$ of Eq.~(\ref{eq_3}) by $k$, $l$ to highlight the new basis. The ground state will be very close to having all three quarks in their respective ground states, namely
$\ket{0}= \ket{\hat{0}\hat{0}\hat{0}}$. 

The first excited state is reasonably represented by elevating exactly one quark to its first excited state (in the philosophy of atomic \emph{Aufbau}), $\ket{\hat{1}}$. Since the colour wavefunction is antisymmetric for a baryon,  the spatial part is here symmetric, so \begin{equation}
\ket{1}= \frac{1}{\sqrt{3}}\Big(\ket{\hat{1}\hat{0}\hat{0}}+\ket{\hat{0}\hat{1}\hat{0}}+\ket{\hat{0}\hat{0}\hat{1}}\Big)\ .
\end{equation}

For the second excited state we can choose between $\ket{\hat{1}\hat{1}\hat{0}}$ and $\ket{\hat{2}\hat{0}\hat{0}}$.
In the harmonic oscillator, with equispaced levels, they would be degenerate. But since the linear potential has slower growth with $r$, 
the second excited state must be naturally closer to $\ket{\hat{2}\hat{0}\hat{0}}$. Of course, configuration mixing can be added in a more accurate treatment.

This is visible in the charmonium potential matrix elements of table~\ref{tabla_8}, where the gap between $V_{00}$ and $V_{11}$ is larger than that between $V_{11}$ and $V_{22}$. Symmetrizing, the variational approximation to the second excited state is of the form \begin{equation}\ket{2}= \frac{1}{\sqrt{3}}\Big(\ket{\hat{2}\hat{0}\hat{0}}+\ket{\hat{0}\hat{2}\hat{0}}+\ket{\hat{0}\hat{0}\hat{2}}\Big)\ .
\end{equation}

We then turn to the terms in the Hamiltonian of Eq.~(\ref{eq_3}), starting by the potential $V_{{k}{l}}\equiv\braket{{k}|V|{l}}$; 
because the potential is two-body, its matrix elements take the form
\begin{equation}
\begin{array}{cc}
&V_{kl}=\braket{n'_1n'_2|V_{12}|n_1n_2}\delta_{n'_3n_3}\\&+\braket{n'_1n'_3|V_{13}|n_1n_3}\delta_{n'_2n_2}+\braket{n'_2n'_3|V_{23}|n_2n_3}\delta_{n'_1n_1}.\label{esta}
\end{array}
\end{equation}
with each term of the form 
\begin{equation}
\begin{array}{cc}
&\braket{n'_in'_j|V_{ij}|n_in_j}=\\&\int_{0}^{\infty} \psi_{n'_i}\psi_{n_i}r_i^2dr_i\int_{0}^{\infty}\psi_{n'_j}\psi_{n_j}r_j^2 dr_j\int_{-1}^{1}V_{ij} d\cos\theta_{ij} \label{eq_24} \ .
\end{array}
\end{equation}
Here,  $\psi_{n_i}$ is the wavefunction of the $i$th quark in the $n$th state as given by Eq.~(\ref{eq_22}). The  integration is straightforward with negligible error.

Likewise, the kinetic energy can be calculated from 
Eq.~(\ref{eq_21.b}), 
\begin{equation}
\begin{array}{cc}
&\braket{k|T|l}= \frac{2}{3}\Bigr(\braket{n'_1|T_1|n_1}\delta_{n'_2n_2}\delta_{n'_3n_3}\\&+\braket{n'_2|T_2|n_2}\delta_{n'_1n_1}\delta_{n'_3n_3}+\braket{n'_3|T_3|n_3}\delta_{n'_1n_1}\delta_{n'_2n_2}\Bigr)\label{eq_27.a}
\end{array}
\end{equation}
The basis $|n\rangle$ does not make $T$ diagonal, but its elements $\braket{n'_i|T_i|n_i}$ are extracted from Eq.~(\ref{T_matrix_element}) and explicitly given in table~\ref{tabla_1.a} (where they can be identified as are proportional to $\omega$). 
\subsection{Numerically computed
$\Omega_{ccc}$ mass \label{sec_3.3}}
Having the complete Hamiltonian at hand we can proceed, with the variational setup of subsection~\ref{subsec:variational}, to measure the mass of the triply charmed baryon $\Omega_{ccc}$.

First, a quick mapping of the variational $d_c$ parameter is performed on a classical simulator running on a standard computer (for economy reasons). 
Minimizing the ground-state $\langle H\rangle$  we obtain the value thereof quoted on table~\ref{tabla_5a} with the numerical results for the potential given in table~\ref{tabla_8}.

 This precomputation produces the values of $\alpha$ and $\beta$ characterizing the ground state ket to be 4.8 and 3.05 radians, respectively.
At this point we are in possession of the ground-state's variational wavefunction, quoted in table~\ref{tabla_10},
which is dominated by the lowest-energy eigenfunction from Eq.~(\ref{Laguerre}).

Then we proceed to recomputing the expetation value of $H$ with IBM's \emph{Osaka} quantum computer, extracting also the $2\sigma$ uncertainty $\Delta (\langle H\rangle)$ corresponding to 95\% confidence level over 10 independent measurements on the Eagle chip.

Finally, to the  energy so obtained we sum the quark masses to mount the baryon mass according to
$M_{ccc}=3m_c+\braket{H}_\psi$, with the charm mass $m_c$  fit on subsection~(\ref{resultadose}) above and quoted in table~\ref{tabla_5a}.

The resulting mass assigned to the $\Omega_{ccc}$ baryon is then $m_{\Omega_{ccc}}=4980\pm 230$ MeV. 
This is in agreement with the earlier pNRQCD computation of 4900$\pm$250MeV at Complutense~\cite{Llanes-Estrada:2011gwu}, and other approaches in table~\ref{tabla_2}.

The uncertainty is totally dominated by our estimate of the error in the kinetic energy (which can be controlled with somewhat more sophisticated few-body treatments). Indeed, the uncertainty on the quantum computer measurement is smaller than 8 MeV at 95\% confidence level.  The uncertainty propagated with the quark mass is about 5 MeV per constituent quark.

\section{Turning to the $b$ quark. \label{sec_4}}

\begin{table*}
\caption{Masses, in MeV, of triply-heavy baryons (all three $b$ or $c$ quarks) computed from Eq.~(\ref{eq_22}), (\ref{esta}) and (\ref{eq_24}), and compared to some standing calculations from the literature. 
The uncertainties include $2\sigma$ from the quantum computer measurement distribution and the uncertainty in the estimate of the cm recoil.
\label{tabla_2}}
\begin{center}
\begin{tabular}{| c | c |c|c|c|c|c|c|}
 \hline
Mass / Particle$_{\rm valence}$ &$\Omega_{bbb}$&$\Omega_{bbc}$&$\Omega_{bcc}$ &$\Omega_{ccc}$\\ \hline \hline
This work & 14270$\pm$190& 11270$\pm$200& 8180$\pm$220& 4980$\pm$230\\\hline
Variational pNRQCD \cite{Llanes-Estrada:2011gwu} & 14700$\pm$300&11400$\pm$300& 8150$\pm$300&4900$\pm$250  \\ \hline
Variational Coulomb method \cite{Yu} &14370$\pm$80&11190$\pm$80&7980$\pm$70& 4760$\pm$60 \\ \hline
QCD sum rules \cite{Z}&13280$\pm$100&10460$\pm$110&7443$\pm$150&4670$\pm$ 150  \\\hline
MIT bag model $M_{+1}$\cite{HASENFRATZ1980401} &14300&11200&8030& 4790 \\\hline
Quark counting rules \cite{Bjorken:1985ei} &14760$\pm$180&11480$\pm$120&8200$\pm$90&4925$\pm$90   \\\hline
Lattice QCD \cite{Brown:2014ena} & 14366$\pm$20&11195$\pm$20& 8007$\pm$20&4796$\pm$18  \\\hline
\end{tabular}
\end{center}
\end{table*}

\subsection{Mesons: $b\bar{b}$ and $b\bar{c}$ \label{4.2}}

In proceeding to baryons containing the $b$ quark we need the mass thereof within the Cornell model. 
We employ the spin-averaged masses of the ground-state bottomonia $b\bar{b}$, experimentally well determined to be, for the $J=0$ $\eta_b$, $(9398.7\pm2.0)$ MeV, and for the $J=1$  $\Upsilon$ $(9460.40\pm0.09)$ MeV, averaging to $\frac{3M_\Upsilon+M_{\eta_b}}{4} = (9445.0\pm0.6)$  MeV.

A straightforward computation yields the model mass of $(9450\pm 11)$ MeV which compares very well with that number and 
gives us a $b$-quark mass of 4594 MeV, with an uncertainty of about 5 MeV.

 Throughout this section and in all that follows, the necessary $\alpha_s$ value is obtained from the running coupling in Eq.~(\ref{runningalpha}) keeping $\Lambda$ constant and adopting as the interaction scale $Q$ the geometric mean of the participating-quark masses, $\sqrt{M_iM_j}$, where $i,j$ are quark flavours, see table~\ref{tabla_9}).  The matrix elements of the potential are then computed and given in table~\ref{tabla_8} of the appendix.

The state $\ket{\psi(\alpha,\beta)}$ for the $b\bar{b}$ ground state meson is obtained from the two-body Hamiltonian whose matrix elements are listed table~\ref{tabla_1}.

The resulting parameters necessary to mount $\psi$ from Eq.~(\ref{eq_16}) come out to be, respectively, $\alpha=1.45$ and $\beta=0.25$ radians (close to the values of 1.45 and 0.15 radians obtained for the $c\bar{c}$ mesons).
Table~\ref{tabla_10} collects the amplitudes of the computational basis elements composing the ground-state wavefunction.

Before proceeding to baryons, we can use the setup so far to predict the mass of the spin-averaged $B_c$ meson as a sanity check:

The experimental mass for its pseudoscalar ground state is $J^P=0^-$  $M_{B_c}=(6274.47\pm0.27)$ MeV, but unfortunately the vector meson has not yet been experimentally discovered, so we are content with obtaining a calculated average somewhat above this pseudoscalar ground state. The coefficients which determine the state of least energy turn out to be $\alpha\simeq  4.85$ and $\beta\simeq 3.35$ radians.  These angles are different from those for the equal-flavoured mesons only apparently:
$\alpha_{bb}\simeq 2\pi-\alpha_{
bc}$ and 
$\beta_{bb}\simeq \beta_{bc}-\pi$, they are so because of different branch choices  for the inverse trigonometric functions (both heavy quarks being a reasonable approximation to an identical static colour source, one expects the angles, as the underlying states, to be reasonably similar).

 Again, the state is spelled out in table~\ref{tabla_10} and the numerical values of the potential which produces it in table~\ref{tabla_8}.

Finally, the resulting meson mass is $M_{B_c}=6255 \pm 12$ MeV which is a bit on the low side of the physical $B_c$ meson, but very reasonable for a purely nonrelativistic Cornell model (see table~\ref{tabla_10} for the ground-state amplitudes and table~\ref{tabla_8} for the potential matrix elements.)

\subsection{All-heavy baryons composed of $b$ (and $c$) quarks  \label{sec_4.3}}

Once the $b$-quark mass has been set in the model, we may predict the mass of the
$\Omega_{bbb}$ baryon, following the same method of section~\ref{sec_3}.
Because $\alpha_s$ is now evaluated at $Q^2=m_b^2$, the values of the potential 
are recomputed and provided in table~\ref{tabla_8}.

We may then repeat sections~\ref{sec_2.2}, \ref{subsec:variational} and \ref{sec.2_4} to extract the minimum-energy state $\ket{\psi(\alpha=4.75,\beta=3.25)}$ (with the angles in radians), collected in table~\ref{tabla_10}. The baryon mass, 
 $M_{bbb}=3m_b+\braket{H}_\psi$, is in turn given in table~\ref{tabla_2}.

With the numerical values of the potential from table~\ref{tabla_1.a} we can once more minimize $\langle H \rangle$ to obtain\\ $\ket{\psi(\alpha=1.65,\beta=3.2)}$, for the $\Omega_{bbc}$ state, and $\alpha= 1.45$, $\beta=6.25$ (all angles in radians) for $\Omega_{bcc}$. This results in a mass of $M=11270\pm200$ MeV, see tables~\ref{tabla_2},~\ref{tabla_10}.

This baryon family is not experimentally known either, 
but several calculations can be found in the literature and we see in table~(\ref{tabla_2}) that our computation is reasonably compatible with the extant predictions in various {\it ab initio} and model approaches, perhaps excepting the sum rule method~\cite{Z} which extracts systematically lower values, as seen in table 12 of~\cite{Llanes-Estrada:2011gwu}.
Again, our quoted uncertainty is dominated by the kinetic energy approximation in one-body coordinates, with a sizeable amount expected from the variational wavefunction, and negligible ones from the quantum computer spread as well as the model parameters. The model systematic uncertainties are more difficult to quantify, of course. Thus, we turn to systems with strange quarks where some experimental input is at hand to help with the systematics.

\section{Baryons  with strangeness \label{sec_5}}
The goal of this section is to combine strange and heavy quarks to obtain
baryons which are starting to appear in experimental efforts, even though the nonrelativistic approximation is invalid for the strange quark (we will lift that approximation too, as the kinetic energy is concerned). Here, approaches based on chiral symmetry for the lighter quark and heavy-spin symmetry for the heavy one have also been worked out~\cite{Romanets:2012hm}. The first order of business is then to fix the strange quark mass. Because the pseudoscalar $0^-$ $s\bar{s}$ state is mixed among $\eta$ and $\eta'$, a notoriously complex system, we prefer to fix the $s$-quark mass with the basic all-strange baryon as follows.

\subsection{Gell-Mann's $\Omega^-$ (sss) \label{sec_5.1}}

We shall extract the strange quark mass from this well-known baryon  via $3M_{s}+\braket{H}_\psi=M_{\Omega_{sss}}=$ 1672.45$\pm$0.29 MeV.  This requires the potentials computed from Eq.~(\ref{eq_22}), (\ref{esta}) and (\ref{eq_24}), followed by the minimization of $H_3$ in the, by now usual, $\ket{\psi(\alpha,\beta)}$ state.
The numerical values of the potentials can be found in table~\ref{tabla_8}. The Coulombic part is now decidedly nonperturbative, with $\alpha_s= 2.02$
which is probably too high (the perturbative running misses the saturation at zero momentum leading to the infrared conformal fixed point).
This yields $m_s=289$ MeV with the ground state having $\alpha$=4.85 and $\beta=$2.6 radians (see table \ref{tabla_10}) and the computed mass being given in the second row of table~\ref{tabla_4}.

Having fixed the strange-quark scale, we may then study baryons composed by combinations of the
 $s$ and the $b$/$c$ quarks, for example  $\Omega_{scc}^+$  and, already experimentally observed,  $\Omega_{ssc}^0$.

\subsection{Mixed  $s+(c,b)$ baryons \label{sec_5.2}}

Employing the potentials of table~\ref{tabla_8} and the matrix elements of table~ \ref{tabla_1.a},  minimization of the Hamiltonian then yields $\left|\psi\left(\alpha=4.85,\beta= 2.70\right)\right\rangle_{ssc}$ and  \\ $\left|\psi\left(\alpha=1.45,\beta= 5.95 \right)\right\rangle_{scc}$, with all angles in radians; the basis decomposition of both $ssc$ and $scc$ states are to be found in table~\ref{tabla_10} and their masses in table~\ref{tabla_3a}.

Our next step is to combine the strange and $b$ quarks, to name the flavour combinations $sbb$, $sbc$ and $ssb$. We follow the same method as for the other baryons. The geometric means of the quark masses and the  $\alpha_s$ constant at their scale are listed in table~\ref{tabla_9}. 
The respective values of the variational parameters  $\alpha$, $\beta$ are  found to be in line with the earlier flavour combinations, namely  4.85, 2.95  for $sbb$; 4.85, 2.70 for $ssb$; and 1.50, 6.05 for $scb$,  (all in radians). Once more, the decomposition of the states $\ket{\psi(\alpha,\beta)}$ can be found  in table~\ref{tabla_10} in the appendix.

The total baryon mass
\begin{equation}
M=M_1+M_2+M_3+\braket{H}_\psi
\label{MountTheMass}
\end{equation}
can now be compared to experiment in the (ssc) flavour combination since  $ \Omega^0_{ssc}$ has been experimentally detected.
Because we are working with central potentials, we need to average the experimental masses of the $3/2$ (mass 2695.2$\pm$1.7 MeV) and $1/2$ (mass 2765.9$\pm$2.0 MeV) spins yielding $\frac{\Omega_{c}^0(J=1/2)+2\cdot \Omega_{c}^0(J=3/2)}{3}=$ 2742.3$\pm$1.4 MeV as experimental reference.

Looking now at  $ \Omega_{b}^- (ssb)$, we 
face the difficulty that the  $J=3/2$ state has not been definitely identified. There seem to be candidate resonances, particularly a triplet around 6340 MeV \cite{Review}, with measured masses (6330$\pm$0.3) MeV, (6340$\pm$0.3) MeV and (6350$\pm$0.4) MeV, plus an isolate  (6316$\pm$0.3) MeV baryon. 
Temptatively, we will adopt as empirical value for  $\Omega_{b}^-(J=3/2)$ a mass of (6330$\pm$20) MeV. Knowing that $\Omega_{b}^-(J=1/2)$ weighs 6045.2$\pm$1.2 MeV, 
the spin average $\Omega_{ssb}^-$ mass  
which we adopt as reference is   6235$\pm$13 MeV.

In table~\ref{tabla_3a} 
we compare our computed strange baryon masses with the experimental one (for the known cases) and some outstanding computations from the literature. Because these references \cite{Shah}, \cite{D}, \cite{Shahs} and \cite{Zhang:2009iya} do treat spin, so that masses depend on the baryon's angular momentum, we have taken spin averages as appropriate.
Generally speaking, our computations treating the $s$ quark as nonrelativistic, with its mass fixed by $\Omega^-_{sss}$, overshoot the accepted masses or mass predictions of the other baryons by quite some.

Therefore, we dedicate an extended subsection to exploring what the cause of this excess energy for strange baryons might be.

\subsection{Systematics of strange-baryon calculation \label{sec_5.3}}

\begin{figure}
  \begin{center}
        \includegraphics[width=\columnwidth]{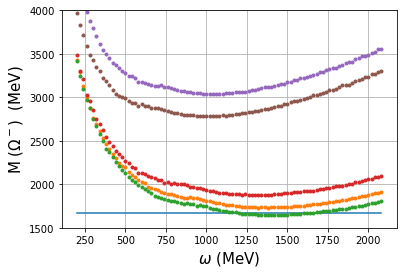}\\
        \includegraphics[width=\columnwidth]{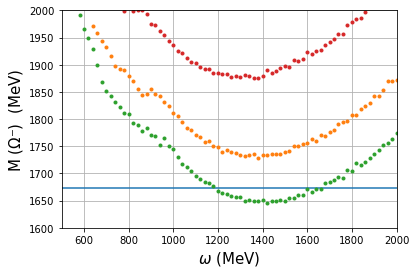}
    \caption{ \label{fig_6} $\Omega^-_{sss}$ mass as function of $\omega$ for various $s$-quark masses (the experimental value is marked by the horizontal line). 
    The upper plot shows, from top to bottom, curves with the parameters
    ($m_s$ (MeV), $\alpha_s$)= (215, 1), (300, 1), (320, 1.56), (310, 1.71) (305, 1.77).
    These values of $\alpha_s$ are as Eq.~(\ref{runningalpha}) dictates, except the first two which were truncated to 1 in the spirit of an IR fixed point.
   The lower plot is a close-up look to the region near the experimental value. 
(The minimization of $\braket{H_3}$ over $\ket{\psi(\alpha,\beta)}$ has not been carried out globally, but only locally, finding two nearby minima at  ($\alpha,\beta$)= (1.4, 6) and (1.5, 0) radians, for speed of computation.)}
    \end{center}
\end{figure}

\paragraph{Running of $\alpha_s$.}
The first idea which we explore is to saturate the infrared behaviour of $\alpha_s$~\cite{alpha,Gonzalez:2011zc,Alkofer:2004it}.
Since the evolution of $\alpha_s$ is computed at one-loop, but it becomes large at low scale as seen in table~\ref{tabla_8} for the $s-s$ interaction, implying nonperturbativity, we explore whether any changes arise from its saturation.
Figure~\ref{fig_6} shows that limiting $\alpha_s$ to 1
and exploring the $m_s$ parameter does not yield any point of agreement with the $\Omega^-$ $(sss)$ mass. 
This seems to come about because diminishing
$m_s$ increases $\braket{V}$ (for Coulombic systems the wavefunction minimizing  $\braket{H_3}_\psi$ has a characteristic radius dropping as $1/m$, and some of this behaviour must remain even with the linear potential added) so that the terms of Eq.~(\ref{MountTheMass}) compensate each other and lead to little improvement.
Thus, the value of $\alpha_s$ is not a leading driver for the strange baryons.

\begin{figure}
    \centering
    \includegraphics[scale=0.60]{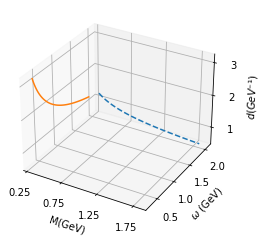}
    \caption{ Two different ways of exploring the variational-wavefunction scale parameter $d$ (in Eq.~\ref{eq_22}) whose
possible values lie on the surface $d=(\omega\ m)^{-0.5}$. 
The simplest is to fix the quark mass of the Hamiltonian and then vary $d$ (continuous line), but we may also keep $\omega$ fixed so that $d$ is automatically determined by $m$ (dashed line).}
    \label{fig:slices}
\end{figure}

\paragraph{Variational scale.}
Another source of systematics is the variational wavefunction in (\ref{eq_22}), particularly when combining different flavours.  
The first aspect thereof that we address is its scale $d$. In figure~\ref{fig:slices} we show two slices that we use to determine possible values of $d$ when given $\omega$ and/or $m$, the kinetic energy parameter and quark mass respectively.

One possibility is to fix $\omega$ which immediately determines $d(m)$ (dashed line, blue online). 
We may also run along the solid line in the figure, so that while the quarks $b$ and $c$ have a fixed $\omega=$562.9 MeV, that for the strange quark is free, so that $\omega(m_s)$ is changing (fixing its value is equivalent to fixing the variational parameter $d$).

The idea is to use this variational freedom to locate a minimum with a better overall spectrum (for example, if $\Omega^-$ would yield a higher $s$ mass, we could have a lower $\alpha_s(m_s)$ and improve the computations of the heavy-strange masses which are on the high side).

The result of the second procedure, with more variational freedom, is naturally better: for strange quark masses in the range 305-310 MeV and $\omega\sim$ 1350-1400 MeV, the experimental $\Omega^-$ is reproduced. 

Therefore, adopting $m_s\simeq$307 MeV in the Hamiltonian and $\omega_s$=1400 MeV in the variational wavefunction, amounting to a scale parameter $d_s=1.53\cdot10^{-3}$, the computed potentials, resulting states, and values of $\alpha_s$ can be read off tables~\ref{tabla_9}, \ref{tabla_8.a},  and~\ref{tabla_10.a} in the appendix, with the resulting masses in table~\ref{tabla_3a} here~
\footnote{ For completeness, the values of  $\alpha$ and $\beta$ for these baryons are now $\Omega^-_ {sss}$: (4.8, 3.15),   $\Omega_{bbs}^-$: (4.6,6.2), $\Omega_{ccs}^+ $: (4.85,3.00), $ \Omega_{bcs}^0 $: (4.6,6.15), $\Omega_{ssb}^-$: (4.65,6.05), and $\Omega_{ssc}^0$ (4.8, 3.0).
 }.

\begin{table*}
\caption{Heavy-strange baryon masses (in MeV). Our computations are shown in the first three numerical rows, with the rest included for comparison. 
The first computation fixes the variational  scale $d(m_i)$ so that $\omega=$562.9 MeV for all quarks. The second however allows the $s$ quark to have a different scale with $\omega_s=1400$ MeV while $\omega_c=\omega_b=$562.9 MeV.  Our last computation takes a different kinematic limit in which the kinetic energy of $s$ is ultrarrelativistic with $b$, $c$ nonrelativistic, showing better agreement with other approaches}.\label{tabla_3a}
\begin{center}
\begin{tabular}{| c | c |c|c|c|c|c|}
 \hline
Mass / Particle$_{\rm valence}$ &$\Omega^-_ {sss}$&$\Omega_{bbs}^-$  &$\Omega_{ccs}^+ $ & $ \Omega_{bcs}^0 $&$\Omega_{ssb}^-$ &$\Omega_{ssc}^0$ \\ \hline
$M$ with flavour-independent $\omega$ &1680$\pm$340&10940$\pm$260&4410$\pm$290 & 7720$\pm$270&  6760$\pm$320& 3320$\pm$320\\ \hline
$M$ with distinguished $\omega_s=1400$ MeV &1700$\pm$530&10920$\pm$550&4400$\pm$540 & 7700$\pm$570&  6800$\pm$960& 3330$\pm$740\\ \hline
Ultrarrelativistic $E_s\simeq |p|$ & 1670$\pm$190&10700 $\pm$270& 4060$\pm$270&7430$\pm$270 &6460$\pm$ 190&2910  $\pm$190\\ \hline \hline
Experimental &1672.45 $\pm$ 0.29 &-  &-&-&6235 $\pm$ 13 &2742.3 $\pm$ 1.4 \\ \hline
  HCQM model \cite{Shah} &-&10460&3757&7170&-&-  \\ \hline
Relativistic quark model \cite{D} &-&10379&3840&7116&-&-  \\ \hline
Hypercentral quark model\cite{Shahs} &-&10332&3636&6964&-&-  \\ \hline
QCD sum rules \cite{Zhang:2009iya} &-&10137&3957 &7367&-&-  \\ \hline
\end{tabular}
\end{center}
\end{table*}

 We see in table~\ref{tabla_3a} that both ways of proceeding, with fixed or sliding $\omega$, yield similar results. 
 We believe this is due to the increase of the kinetic energy of the $s$ quark upon increasing $\omega$. This overestimates the resulting baryon mass.

\paragraph{Kinetic energy in one-body, nonrelativistic approximation.}
The largest source of error with the present setup is doubtlessly the treatment of the kinetic energy of the center of mass, especially when we mix light and heavy quarks. The crude treatment in one-body coordinates (neglecting crossed-terms in the center of mass Laplacian) means that we have probably systematically undersubtracted the CM energy, so the masses involving the strange quark, which is more sensitive to this, are systematically higher.

It is also necessary to keep in mind that the $s$-quark is less amenable to a nonrelativistic treatment. This is caused by the fact that the Lorentz time dilation factor $ \gamma =\sqrt{m^2+p^2}/m_s \simeq \braket{\frac{T}{3}}/m_{s}+1 = 2.57$ (for a flavor-independent $\omega$) or 3.28 (for the strange quark distinguished with $\omega_s=1400MeV$)  which points to a  substantially relativistic system.
We could continue adding terms in the expansion in powers of $p/m$ (but two at a time need to be added~\cite{Lucha:2014jca} to guarantee a Hamiltonian bound from below, and this complicates the calculation). 
In principle, a more relativistic treatment of baryons even in the valence-quark approximation is possible~\cite{Ebert:2017veh}, but we believe it unnecessary for our thrust in exploring a small qubit-number based calculation.

It is useful to look at a computation in the opposite, ultrarrelativistic limit (for the $s$-quark only), that is, instead of expanding $\sqrt{m^2+p^2}\simeq m + p^2/(2m) + \dots$, employing 
\begin{equation}
    \sqrt{m^2+{\bf p}^2}\simeq |{\bf p}| + \frac{m^2}{2|{\bf p}|}+\dots \label{URexpansion}
\end{equation}

The outcome of such calculation, with nonrelativistic heavy quarks and an ultrarrelativistic kinetic energy for the strange one, is also given in table~\ref{tabla_3a}. The improvement is noticeable. 

Unlike the earlier difficulties with the nonrelativistic expansion (Fig.~\ref{fig_6}), this relativistic setup, with $\alpha_s\to 1$ saturated in the infrared,
easily fits the $\Omega_{sss}$ baryon, and the strange quark mass turns out to be $m_s\simeq$ 188 MeV with variational wavefunction parameter $d_s= 3.07 \cdot 10^{-3}$. 

This modification requires one new matrix element, 
\begin{equation}
\braket{p_j}=\braket{n'_j|p_j|n_j}=\int_0^\infty r^2_j\psi_{n'_j}\partial_{r_j} \psi_{n_j} .
\end{equation}

Then, the second term in the ultarrelativistic expansion of Eq.~(\ref{URexpansion}) for the $j$th quark is approximated as follows
\begin{equation}
    E_j\simeq \braket{T_j}= \braket{n'_j|T|n_j} \simeq  \braket{p_j} + \frac{m^2}{2\braket{p_j}} \label{Trelativistic}\ .
\end{equation}
The time-dilation factor for a strange quark in the $\Omega_{sss}$ baryon is then $\gamma\simeq2.35$.

The matrix elements of the Hamiltonian are also quoted in the appendix. The kinetic energy $\braket{n'_j|T|n_j}$ from Eq.~(\ref{Trelativistic}) for the ultrarrelativistic $s$-quark, and still employing Eq.~(\ref{T_matrix_element}) for the nonrelativistic $c$, $b$ ones.

The respective wavefunction parameter values for
$\alpha$ and $\beta$ fed into the quantum computer are $\Omega^-_ {sss}$ (4.55, 5.55),   $\Omega_{bbs}^-$ (1.65,2.55), $\Omega_{ccs}^+ $ (1.7,2.55), $ \Omega_{bcs}^0 $ (1.65,2.55), $\Omega_{ssb}^-$ (4.85,2.35), $\Omega_{ssc}^0$ (4.55, 5.6).

Another source of systematic error is the wavefunction size, $N$ in Eq.~(\ref{eq_3}).
The sign of the discrepancy with data is here known to be positive, as guaranteed by the variational theorem (the computed ground state mass with a given Hamiltonian $H$ is an upper bound to the exact value for that same $H$). 

Additionally, when we defined the three-particle basis (subsection~\ref{sec_3.2})
we adopted three equal wavefunctions, which is reasonable for $c$ and $b$, both close to being static sources. We expect this to require modification for the $s$ quark.
We could think of a redefinition in which the first excited state is always 
 $\ket{1}$ for the single $s$ quark (or the appropriately symmetrized combination for baryons with $ss+c/b$ compostion ) while keeping the heavy quark(s) at $\ket{0}$.  Then the second excitation would correspond to setting the heavy quarks in $\ket{1}$ while keeping the strange one in $\ket{0}$. 
 
Increasing the complexity of the variational wavefunction $\ket{\psi(\alpha,\beta)}$ exceeds the ambition of this article and we leave it to future work.

\section{Conclusion}
We have studied the basic spectrum of 
baryons in the valence Cornell quark model in any combination of $c$, $b$ or $s$ quarks, extending the model approach of Gallimore and Liao.

The masses of the triply-heavy baryons were presented in table~\ref{tabla_2}, 
and they are in line with other predictions in the literature.

Because the strange quark is on the light side of the QCD mass scale, the baryon spectrum including them (sections \ref{sec_5.2} and \ref{sec_5.3}) escapes the nonrelativistic treatment and requires further work. We have deployed far more sophisticated models of baryons in the past~\cite{Bicudo:2009cr,Bicudo:2016eeu} which, in addition to elevating the Cornell potential to a field theory, include fully relativistic spinors and spontaneous chiral symmetry breaking to be able to treat light quarks; but model sophistication is not the point here, rather to explore this new computational method based on digital quantum computers. We have also left out strange-light baryons~\cite{Chao:1980em,Estevez:2020vsm} which have very rich physics and deserve separate treatment.

The quantum computer is here used as a small diagonalizer, and there is no quantum advantage for these small few-body systems. But the note of interest is that present machines already induce sufficiently small statistical errors so that our uncertainties are dominated by more traditional sources. These are, saliently, the small variational space here considered, the recoil of the center of mass, and the uncertainty induced by the determination of the model parameters.

It is clear to us that quantum computing is quickly advancing and will soon be an additional tool in the box of practicing nuclear and particle physicists, with interesting possibilities opening up.

\newpage

\bibliographystyle{spphys}
\bibliography{porfi}

\newpage
\section*{Appendix: Tabulated numerical parameters and computed matrix elements of the Hamiltonian.}
Here we collect a number of tables containing numerical data, parameter values, and expressions for terms of the 2- and 3-body Hamiltonians necessary for full reproductibility of the spectra computed in this work. Also given are the states $|\psi(\alpha,\beta)\rangle$ that, within the variational family, minimize the expected energy. 

Most of the tables are self-explanatory, but a comment on the kinetic energy in table~(\ref{tabla_1.a}) is in order. 
Expressions of the type $\braket{k'|T|k}$ in this table all have equal $\omega$ parameter. If either $k$ or $k'$ corresponds to an excited state of the oscillator, either of the quarks can have been promoted to it. Then, $\braket{k'|T|k}$ is to be understood as averaging over which of the quarks occupies the higher state. Individual components of that matrix element can be either relativistic of Eq.~(\ref{Trelativistic}) or nonrelativistic of~Eq.(\ref{T_matrix_element})
for example, if we have only one $s$-quark, 
 \begin{equation}
     \braket{1|T|0}=\frac{1}{3} \braket{1|T|0}_{\rm relativistic}+\frac{2}{3} \braket{1|T|0}_{\rm nonrelativistic}\ .
 \end{equation}  
\begin{table*}
\caption{Numerical values of constants appearing in 
the $c\bar{c}$ computation of section~\ref{sec:charmonium}. \label{tabla_4}  }
\begin{center}
\begin{tabular}{| c | c |c|c|c|c|c|}
\hline
Constants & $\alpha_s$ &$\sqrt{\sigma}$ &$\mu $&$\omega $&$d $  \\ \hline
Values&0.4038 &441.6 MeV&645 MeV&562.9 MeV&1.17$\cdot10^{-3}$ MeV$^{-1}$ \\ \hline
\end{tabular}
\end{center}
\end{table*}

\begin{table*}
\caption{Values of the quark-flavour dependent quantities. \label{tabla_5a}}
\begin{center}
\begin{tabular}{| c | c |c|c|}
\hline
Quark & $c$ &$b$&$s$ \\ \hline
Mass (MeV) & 1290 & 4594 & 289 \\ \hline
$d_i$ baryon& 1.17$\cdot10^{-3} $ &
6.22$\cdot10^{-4} $&2.48$\cdot10^{-3} $\\ 
constants  & & & \\
\hline
\end{tabular}
\end{center}
\end{table*}

\begin{table*}
\caption{Matrix elements of the $q\bar{q}$-meson Hamiltonian in terms of Pauli matrices. First computed in  \cite{gallimore-2023}. \label{tabla_1} }
\begin{center}
\begin{tabular}{| c | c |c|c|}
\hline
H$_3^0$  & $\frac{1}{2}(\frac{21}{4}\omega+V_{00}+V_{11}+V_{22}) \mathbbm{1}_{2\times2}$ & $H_3^5 $ & $\frac{1}{4}(-\sqrt{5}\omega+2V_{12})\sigma^x_1\sigma^x_2$  \\ \hline
H$_3^1 $ & $-\frac{1}{2}(\frac{3}{4}\omega+V_{00})\sigma^z_0$ & $H_3^6 $ & $\frac{1}{4}(-\sqrt{\frac{3}{2}}\omega+2V_{01})\sigma^y_0\sigma^y_1$\\ \hline
H$_3^2 $ & $-\frac{1}{2}(\frac{7}{4}\omega+V_{11})\sigma^z_1$& $H_3^7 $ & $\frac{1}{4}(-\sqrt{5}\omega+2V_{12})\sigma^y_1\sigma^y_2$\\ \hline
$H_3^3$ & $-\frac{1}{2}(\frac{11}{4}\omega+V_{22})\sigma^z_2$& $H_3^8 $ & $\frac{1}{2}V_{02}\sigma^x_0\sigma^z_1\sigma^x_2$ \\ \hline
$H_3^4$ & $\frac{1}{4}(-\sqrt{\frac{3}{2}}\omega+2V_{01})\sigma^x_0\sigma^x_1$ & $H_3^9 $ & $\frac{1}{2}V_{02}\sigma^y_0\sigma^z_1\sigma^y_2$\\ \hline
\end{tabular}
\end{center}
\end{table*}

\begin{table*}
\caption{For a system of three quarks with identical kinetic energy parameter $\omega$, values of the Hamiltonian matrix elements. \label{tabla_1.a}}
\begin{center}
\begin{tabular}{| c | c |c|c|}
\hline
H$_3^0$  & $\frac{1}{2}(\frac{13}{2}\omega+V_{00}+V_{11}+V_{22})$ $\mathbbm{1}_{2\times2}$ & $H_3^5 $ & $\frac{1}{4}(-2\frac{\sqrt{5}}{3}\omega+2V_{12})\sigma^x_1 \sigma^x_2$  \\ \hline
H$_3^1 $ & $-\frac{1}{2}(\frac{3}{2}\omega+V_{00})\sigma^z_0$ & $H_3^6 $ & $\frac{1}{4}(-\sqrt{{2}}\omega+2V_{01})\sigma^y_0 \sigma^y_1$\\ \hline
H$_3^2 $ & $-\frac{1}{2}(\frac{13}{6}\omega+V_{11})\sigma^z_1$& $H_3^7 $ & $\frac{1}{4}(-2\frac{\sqrt{5}}{3}\omega+2V_{12})\sigma^y_1\sigma^y_2$\\ \hline
$H_3^3$ & $-\frac{1}{2}(\frac{17}{6}\omega+V_{22})\sigma^z_2$& $H_3^8 $ & $\frac{1}{2}V_{02}\sigma^x_0 \sigma^z_1 \sigma^x_2$ \\ \hline
$H_3^4$ & $\frac{1}{4}(-\sqrt{{2}}\omega+2V_{01})\sigma^x_0 \sigma^x_1$ & $H_3^9 $ & $\frac{1}{2}V_{02} \sigma^y_0 \sigma^z_1 \sigma^y_2$\\ \hline
\end{tabular}
\end{center}
\end{table*}

\begin{table*}
\caption{Potential matrix elements in MeV, for the various flavour combinations. Those for  the three-quark system are computed from Eq.~(\ref{eq_22}), (\ref{esta}) and (\ref{eq_24}); and those for the two-quark mesons from Eq.~(\ref{H}, \ref{eq_5} and \ref{eq_6}). \label{tabla_8}}
\begin{center}
\resizebox{17cm}{!} {
\begin{tabular}{| c | c |c|c|c|c|c|c|c | c |c|c|c|c|c|c|c|c|}
\hline
Flavour  quark&$c\bar{c}$&$ccc$ &$b\bar{b}$&$b\bar{c}$&$bbb$&$bcc$& $bbc$& $sss$& $ssc$& $scc$&$ssb$ &$sbb$& $sbc$\\ \hline
V$_{00}$ &90.67&272.00&-114.82&-15.57&-344.80&163.10&-42.50&355.00&771.38&743.72&863.86&630.74 &735.49\\ \hline
V$_{01}$&261.16&452.37&204.88&223.41&355.02&409.31&376.85&1007.4&750.97&565.94&690.41&472.93&514.13\\ \hline
V$_{02}$&-108.51&-94.02&-102.10&-101.34&-88.35&-104.88&-102.99&-215.92&-177.98&-137.35&-177.98&-135,45&-142.78\\ \hline
V$_{11}$&319.02&500.29&33.32&164.25&-196.40&365.05&132.83&849.65&1168.24&1051.79&1237.28&888.61&1018.70\\ \hline
V$_{12}$&316.68&396.82&234.57&263.97&304.18&343.18&312.31&880.52&635.13&473.89&567.49&375.36&413.26\\ \hline
V$_{22}$&481.12&791.33&134.15&289.73&14.61&611.54&352.63&1490.01&1629.18&1396.28&1630.39&1138.60&1306.99\\ \hline
\end{tabular}
}

\end{center}
\end{table*}

\begin{table*}
\caption{ Decomposition of the  $\ket{\psi(\alpha,\beta)}$ (spin-averaged) ground states of the mesons and baryons indicated, obtained upon finding the minimum of the energy in the ($\alpha$,$\beta$) variational space.\label{tabla_10} }
\begin{center}
\resizebox{17cm}{!} {
\begin{tabular}{| c | c |c|c|c|c|c|c|c|c|c|c|c|c|c|c|}
 \hline
Component&
$c\bar{c}_{1s}$&$c\bar{c}_{2s}$&$\Omega_{ccc} $&$b\bar{b}$&$b\bar{c}$&$\Omega_{bbb}$&$\Omega_{bbc}$&$\Omega_{bcc}$ &$\Omega (sss)$&$\Omega_{c}^+ (css)$&$\Omega_{cc}^+ (ccs)$&$\Omega_{bb}^- (bbs)$& $\Omega_{b}^- (ssb)$&$\Omega_{cb}^0 (scb)$\\ \hline
$\ket{100}$ &0.982&-0.178&0.992&0.962&0.969&0.994&-0.995&0.992&0.848&0.886&0.938&0.917&0.896&0.970\\ \hline
$\ket{010}$&0.148&0.942&-0.091&0.246&0.204&0.108&-0.058&-0.033&-0.511&-0.423&-0.324&0.396&-0.423&-0.230\\ \hline
$\ket{001}$&0.121&0.284&0.087&0.121&0.137&0.037&-0.079&0.121&0.137&0.137&0.121&0.037&0.137&0.071\\ \hline
\end{tabular}
}
\end{center}
\end{table*}

\begin{table*}
\caption{Values of the $\alpha_s$ constant employed. 
In the first two sets, the $s$ quark is treated on equal footing with the heavy quarks, in a nonrelativistic expansion. For each quark pair as indicated, we take the scale to be the geometric average of the two masses. These two sets are distinguished by the strange quark mass, the top one with $m_s=289$ MeV, the second one with $m_s= 307$ MeV, respectively. In both, $\alpha_s$ results from the one-loop formula.   
The third set at the bottom of the table corresponds to an ultrarrelativistic treatment of the $s$ quark instead, 
and $\alpha_s$ is taken to saturate in the infrared at a maximum value of 1 (modest variations on top of this number can be absorbed by the strange quark mass if needed, as much as the spectrum is concerned).
\label{tabla_9}}
\begin{center}
\begin{tabular}{| c | c |c|c|c|c|c|}
 \hline
Pair                          &  $c$-$c$ &  $b$-$b$  & $b$-$c$   &  $b$-$s$  &  $c$-$s$  &  $s$-$s$ \\ \hline \hline
$\mu_{12}=\sqrt{m_1m_2}$ (MeV)&    1290  &   4594    &  2434     &   1152    &    611    &   289    \\ \hline
$\alpha_s(\mu_{12})$          &   0.40   &   0.24    &   0.30    &     0.43  &     0.67  &2.03  \\ \hline\hline
$\mu_{12}=\sqrt{m_1m_2}$ (MeV)&          &           &           &   1188    &    629    &307    \\ \hline
$\alpha_s(\mu_{12})$          &          &           &           &    0.42   &    0.66   &1.75  \\ \hline\hline
$\mu_{12}=\sqrt{m_1m_2}$ (MeV)&          &           &           &     929    &   492   &188    \\ \hline
 $\alpha_s(\mu_{12})$        &          &           &           &    0.49   &    0.83  & IR truncated to $\alpha_s=1$  \\ \hline
\end{tabular}
\end{center}
\end{table*}

\begin{table*}
\caption{Numerical value of the potential matrix elements, in MeV, for different quark configurations. 
The valence strange-quark mass is fixed at 307MeV and the characteristic kinetic energyis $\omega_s=1400$ MeV, as obtained from Eq.~(\ref{eq_22}), (\ref{esta}) and (\ref{eq_24}) for three-quark systems. \label{tabla_8.a}}
\begin{center}
\begin{tabular}{| c | c |c|c|c|c|c|c|}
\hline
Flavour  quark& V$_{00}$ &V$_{01}$&V$_{02}$&V$_{11}$&V$_{12}$&V$_{22}$  \\ \hline
 $sss$ &-1315.63&981.40&-254.13&-926.11&837.14&-354.28 \\ \hline
 $ssc$ &206.84& 697.84&-180.39&547.33&583.78&962.22 \\ \hline
 $scc$ &543.35&547.89&-136.97&831.77&457.75&1162.42\\ \hline
 $ssb$ &655.31&646.52&-172.80& 996.49&528.61&1359.68\\ \hline
 $sbb$ &548.47&463.36&-134.40&797.45&367.48&1041.80\\ \hline
 $sbc$ &627.40&501.77&-141.56&898.78&402.88&1179.23\\ \hline
\end{tabular}
\end{center}
\end{table*}

\begin{table*}
\caption{
Basis decomposition of minimum-energy variational states  $\ket{\psi(\alpha,\beta)}$ 
for several baryons including a valence $s$ quark of mass  307 MeV with characteristic kinetic energy constant $\omega_s=1400$ MeV. In all cases the $|100\rangle$ is seen to be dominant. \label{tabla_10.a}}
\begin{center}
\begin{tabular}{| c | c |c|c|c|c|c|c|c|c|c|c|}
 \hline
State&$\Omega^- (sss)$ &$\Omega_{c}^0 (css)$&$\Omega_{cc}^+ (ccs)$&$\Omega_{bb}^- (bbs)$&$\Omega_{b}^- (ssb)$&$\Omega_{bc}^0 (scb)$ \\ \hline
$\ket{100}$&0.996&0.986&0.980&-0.990&-0.971&-0.985\\ \hline
$\ket{010}$&0.008&-0.141&-0.140&0.083&0.231&0.132\\ \hline
$\ket{001}$&0.087&0.087&0.137&-0.112&-0.062&-0.112\\ \hline
\end{tabular}
\end{center}
\end{table*}

\begin{table*}
\caption{ Matrix elements of the Hamiltonian with quarks in different kinematic regimes (namely, an ultrarrelativistic $s$ quark and nonrelativistic $c$, $b$ quarks).   \label{tabla_1.a}}
\begin{center}
\resizebox{14cm}{!} {
\begin{tabular}{| c | c |c|c|}
\hline
H$_3^0$  & $\frac{1}{2}(\frac{2}{3}(\braket{1|T|1}+7\braket{0|T|0}+\braket{2|T|2}))+V_{00}+V_{11}+V_{22})$ $\mathbbm{1}_{2\times2}$ & $H_3^5 $ & $\frac{1}{4}\left(\frac{2}{3}(\braket{1|T|2}+\braket{2|T|1})+2V_{01} \right) \sigma^x_1 \sigma^x_2$   \\ \hline
H$_3^1 $ & $-\frac{1}{2}(2\braket{0|T|0}+V_{00})\sigma^z_0$ & $H_3^6 $ & $\frac{1}{4}\left(\frac{2}{\sqrt{3}}(\braket{0|T|1}+\braket{1|T|0})+2V_{01} \right) \sigma^y_0 \sigma^y_1$ \\ \hline
H$_3^2 $ & $-\frac{1}{2}(\frac{2}{3}(\braket{1|T|1}+2\braket{0|T|0})+V_{11})\sigma^z_1$& $H_3^7 $ &  $\frac{1}{4}\left(\frac{2}{3}(\braket{1|T|2}+\braket{2|T|1})+2V_{01} \right) \sigma^y_1 \sigma^y_2$ \\ \hline
$H_3^3$ & $-\frac{1}{2}(\frac{2}{3}(\braket{2|T|2}+2\braket{0|T|0})+V_{22})\sigma^z_2$& $H_3^8 $ & $\frac{1}{4}\left(\frac{2}{\sqrt{3}}(\braket{0|T|2}+\braket{2|T|0})+2V_{02} \right) \sigma^x_0 \sigma^z_1 \sigma^x_2$ \\ \hline
$H_3^4$ & $\frac{1}{4}\left(\frac{2}{\sqrt{3}}(\braket{0|T|1}+\braket{1|T|0})+2V_{01} \right) \sigma^x_0 \sigma^x_1$ & $H_3^9 $ & $\frac{1}{4}\left(\frac{2}{\sqrt{3}}(\braket{0|T|2}+\braket{2|T|0})+2V_{01} \right) \sigma^y_0 \sigma^z_1 \sigma^y_2$\\ \hline
\end{tabular}
}
\end{center}
\end{table*}

\begin{table*}
\caption{ Numeric values of the potential matrix elements in MeV, analogous to those in table~\ref{tabla_8.a} but for the strange quark treated in the ultrarrelativistic limit, with $\omega_s=1562.9$ MeV for all flavours and $m_s$=188 MeV. \label{tabla_13}}
\begin{center}
\begin{tabular}{| c | c |c|c|c|c|c|c|}
\hline

Quark flavour & V$_{00}$ &V$_{01}$&V$_{02}$&V$_{11}$&V$_{12}$&V$_{22}$  \\ \hline
 $sss$ &2091.15&859.95&-137.18&2612.70&773.82&3215.19\\ \hline
 $ssc$ &1573.36&762.73&-175.43&2020.32&640.07&2489.71 \\ \hline
 $scc$ &966.97&626.87&-161.05&1316.18&514.40& 1681.76\\ \hline
 $ssb$ &1724.871&695.05&-164.49& 2148.34&572.74&2553.31\\ \hline
 $sbb$ &912.89&526.73&-148.20&1211.98&416.18&1486.46\\ \hline
 $sbc$ &988.28&571.50&-161.01&1312.57&453.93&1623.66\\ \hline
\end{tabular}
\end{center}
\end{table*}

\begin{table*}
\caption{ Matrix elements ($\braket{n_j'|T|n_j}$) in MeV of the kinetic energy for the  $s$ quark treated in the ultrarrelativistic limit. \label{tabla_14}}
\begin{center}
\begin{tabular}{| c | c |c|c|c|c|c|c|}
\hline
Relativistic kinetic energy& T$_{00}$ &T$_{01}$&T$_{02}$&T$_{11}$&T$_{12}$&T$_{22}$  \\ \hline
 $s$ &415.21&482.15&548.34&373.79&581.30&332.88\\ \hline

\end{tabular}
\end{center}
\end{table*}

\begin{table*}
\caption{ Basis decomposition of the 
heavy-strange baryon states, analogous to table~\ref{tabla_10.a} but for the ultrarrelativistic kinetic energy of Eq.~(\ref{URexpansion}) with the parameters given in the text.
It is of note that the ket $\ket{100}$ is no more dominant: to relax the wavefunction a second state is necessary. We do not believe this leads to a loss of precision for the ground state, having enough flexibility with three variational wavefunctions and one free variational parameter, but a reasonable computation of excited states probably needs a larger space.  \label{tabla_14.b}}
\begin{center}
\begin{tabular}{| c | c |c|c|c|c|c|c|c|c|c|c|}
 \hline
State&$\Omega^- (sss)$ &$\Omega_{c}^0 (css)$&$\Omega_{cc}^+ (ccs)$&$\Omega_{bb}^- (bbs)$&$\Omega_{b}^- (ssb)$&$\Omega_{bc}^0 (scb)$ \\ \hline
$\ket{100}$&-0.733&-0.765&-0.823&-0.827&0.696&-0.827\\ \hline
$\ket{010}$&0.660&0.623&0.553&0.556&-0.705&0.556\\ \hline
$\ket{001}$&-0.162&-0.162&-0.129&-0.079&0.137&-0.079\\ \hline
\end{tabular}
\end{center}
\end{table*}

\end{document}